\documentclass[iop]{emulateapj}

\def \solar{_\odot}
\def \pow10#1{\times 10^{#1}}

\tighten
\begin{document}

\submitted{Accepted for publication in ApJ}

\title{The faint stellar halos of massive red galaxies from stacks of
  more than 42000 SDSS LRG images}

\author{Tomer Tal\altaffilmark{1},
  Pieter G. van Dokkum\altaffilmark{1}}
\altaffiltext{1}{Yale University Astronomy Department, P.O. Box
  208101, New Haven, CT 06520-8101 USA}

\shorttitle{LRG image stacking}
\shortauthors{Tal \& van Dokkum}

\begin{abstract}
  We study the properties of massive galaxies at an average redshift
  of $z\sim 0.34$ through stacking more than 42000 images of Luminous
  Red Galaxies from the Sloan Digital Sky Survey.
  This is the largest dataset ever used for such an analysis and it
  allows us to explore the outskirts of massive red galaxies at
  unprecedented physical scales.
  Our image stacks extend farther than 400 kpc, where the r-band
  profile surface brightness reaches 30 mag arcsec$^{-2}$.
  This analysis confirms that the stellar bodies of luminous red
  galaxies follow a simple S\'{e}rsic profile out to 100 kpc.
  At larger radii the profiles deviate from the best-fit S\'{e}rsic
  models and exhibit extra light in the g, r, i and z-band stacks.
  This excess light can probably be attributed to unresolved
  intragroup or intracluster light or a change in the light profile
  itself. We further show that standard analyses of SDSS-depth images
  typically miss 20\% of the total stellar light and underestimate the
  size of LRGs by 10\% compared to our best fit r-band S\'{e}rsic
  model of n=5.5 and r$_e$=13.1 kpc.
  If the excess light at r$>$100 kpc is considered to be part of the
  galaxy, the best fit r-band S\'{e}rsic parameters are n=5.8 and
  r$_e$=13.6 kpc.
  In addition we study the radially dependent stack ellipticity and
  find an increase with radius from $\epsilon$=0.25 at $r$=10 kpc to
  $\epsilon$=0.3 at $r$=100 kpc.
  This provides support that the stellar light that we trace out to at
  least 100 kpc is physically associated with the galaxies themselves
  and may confirm that the halos of individual LRGs have higher
  ellipticities than their central parts.
  Lastly we show that the broadband color gradients of the stacked
  images are flat beyond roughly 40 kpc, suggesting that the stellar
  populations do not vary significantly with radius in the outer parts
  of massive ellipticals.
\end{abstract}

\section {\label{intro}Introduction}
 Elliptical galaxies dominate the galaxy mass function for
 $M_{\star}\geq10^{11}M\solar$ and are the brightest extended
 objects in the nearby Universe. 
 Extensive effort has been made to study nearby ellipticals, relying
 on observations of their morphology, kinematics and stellar
 populations \citep[e.g.,][]{faber_velocity_1976,
 illingworth_rotation_1977, boroson_color_1983,
 dressler_spectroscopy_1987, kormendy_surface_1989,
 peletier_elliptical_1989, worthey_mg_1992}.
 Most studies have focused on the bright centers of these galaxies,
 out to 1-2 effective radii, as detection of their faint outskirts has
 posed a challenge both observationally and analytically.
 Nevertheless, there exists great interest in correctly analyzing
 the full physical extent of massive red galaxies, particularly as
 they may build up inside-out through mergers
 \citep{loeb_cosmological_2003, naab_formation_2007,
 bezanson_relation_2009}.
 An accurate measurement of the light profile shape of low redshift
 galaxies is crucial for the interpretation of high redshift galaxy
 observations.

 Direct observations of the outskirts of individual massive
 galaxies are hard to perform given the extremely faint surface
 brightness level that is reached at large radii.
 The observed light in such data is highly influenced by nearby
 objects, flat fielding errors and the wings of the PSF
 \citep[e.g.,][]{mihos_diffuse_2005, de_jong_point_2008,
 tal_frequency_2009, van_dokkum_newfirm_2009}.
 A recent study by \cite{kormendy_structure_2009}
 utilized a compilation of data from several sources including
 observations of 28 Virgo ellipticals.
 The authors detected light outside of five effective radii in only
 two galaxies, M87 and M49, which are the most luminous Virgo
 galaxies.
 There have also been many studies of extended light distributions in
 massive clusters, and in those environments the light at large
 distances from the central galaxy is usually attributed to an
 intracluster stellar population distinct from the central galaxy
 (Intra-Cluster Light, or ICL).
 The ICL has been successfully traced to distances of $>$500 kpc in
 several studies \citep[e.g.,][]{gonzalez_measuring_2000,
 krick_diffuse_2007, mihos_diffuse_2005}.
 
 As an alternative to deep observations, stacking a large number of
 images of similar galaxies could greatly improve the reached overall
 depth at the cost of losing system specific information.
 This technique has been used by several authors in studies of
 both spiral and elliptical galaxies
 \citep[e.g.,][]{zibetti_haloes_2004, van_dokkum_growth_2010}.
 \citet[hereafter Z05]{zibetti_intergalactic_2005} stacked 683 images
 of clusters in SDSS, and reported that the ICL accounts for 10\% of
 the total light in galaxy clusters.

 In this study we stack an unprecedentedly large sample of LRG images,
 in a similar way as Z05, to explore the faint outskirts of massive
 red galaxies.
 LRGs are thought to be mostly group centrals that live in lower mass
 halos than the objects studied in Z05.
 By stacking the LRG data we hope to shed new light on the properties
 of these objects at very large radii and to better constrain the size
 and total luminosity of elliptical galaxies at $z\sim0.34$.

\section{LRG image stacking}\label{secstack}
 \subsection{Sample selection}
  We selected galaxy images for this study from the Sloan Digital Sky
  Survey \citep[SDSS,][]{abazajian_seventh_2009} including all objects
  classified as Luminous Red Galaxies (LRG) that have a spectroscopic
  redshift measurement.
  LRGs are intrinsically red and luminous objects that were identified
  as such from their central surface brightness and location on a
  rotated color-color diagram \citep[for full details
  see][]{eisenstein_spectroscopic_2001}.
  This selection is aimed at finding the most luminous red galaxies in
  the nearby Universe ($L \geq 3L^{\star}$) out to a redshift of
  $z=0.5$.
  Being some of the most massive galaxies in SDSS, LRGs
  occupy the high end of the stellar mass spectrum between
  $10^{11}M_{\solar}$ and a few times $10^{12} M_{\solar}$.
  Roughly 90\% of all LRGs are central halo galaxies and they mainly
  reside in groups with a typical halo mass of a few times $10^{13}
  M_{\solar}$ \citep{wake_2df-sdss_2008, zheng_halo_2009,
  reid_constraining_2009}.

  The seventh data release of SDSS (DR7) includes 188366
  spectroscopic LRGs over a wide range of redshifts and apparent
  magnitudes.
  The redshift distribution has two main populations peaking at
  $z\sim$0.05 and $z\sim$0.34.
  The lower redshift LRG candidates are predominantly a contamination
  sub-sample of fainter, lower mass red galaxies.
  In order to assure that our sample is indeed composed of high mass,
  luminous red galaxies we selected objects with a narrow distribution
  of redshifts around the high-$z$ peak.
  This selection also ensures that galaxies in this distance and
  brightness ranges do not suffer from significant size and mass
  evolution within the sample.

  \begin{figure}
    \includegraphics[width=0.47\textwidth]{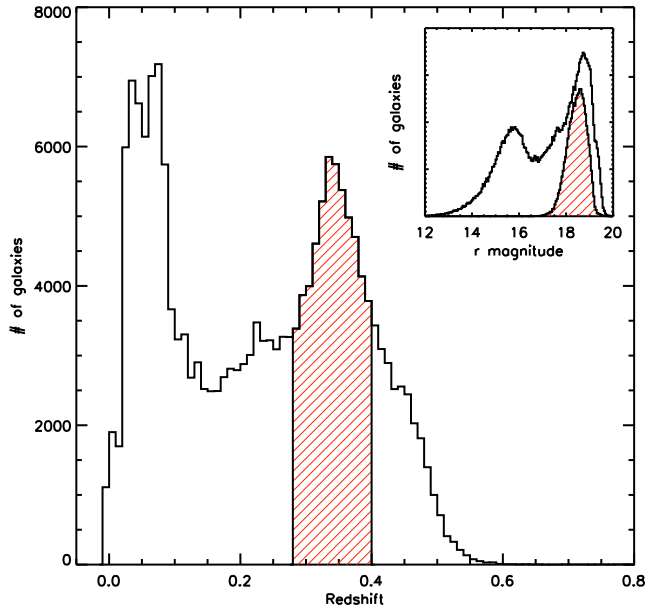}
    \caption{The redshift distribution of the entire LRG candidate
    sample (black contour) overlaid with the cuts applied in
    this study (red area).  The mean value of the selected
    sample is $z\sim$0.34.  The resultant r-magnitude
    distribution is also shown, along with the overall LRG r-magnitude
    distribution, in the top-right corner.}
    \hfill
    \label{fig:zdist}
  \end{figure}

  The main sample used in this study comprises 55650
  galaxies in a redshift range 0.28$\leq$$z$$\leq$0.40 with mean redshift
  $<$$z$$>$=0.34, resulting in an apparent r-magnitude range of
  18.5$\pm$0.4 (figure \ref{fig:zdist}).
  In fact, 99\% of the sample falls within the flux limit of cut I, as
  defined by \cite{eisenstein_spectroscopic_2001}, making it
  approximately volume-limited.
  From this master list we excluded 12750 galaxies (23\%) where more
  than 75\% of the central 5''$\times$5'' had to be masked out due
  to close proximity to another object (masking details in
  subsection \ref{prep}).
  In addition, we excluded 293 ($<$1\%) of the LRGs because of varying
  sky levels in the frame caused by close proximity to a bright star
  in or just outside of the field.
  Finally, we excluded 28 ($<$0.1\%) images of galaxies with apparent
  r-magnitude outside of the selected range (details in subsection
  \ref{magstack}).
  The final sample consists of 42579 galaxies.
  
 \subsection{Preparing the images for stacking \label{prep}}
  We acquired imaging data for the fields containing the selected
  galaxies from the SDSS archive in all five bands: u, g, r, i and z,
  corresponding to central wavelengths of 355.1$nm$, 468.6$nm$,
  616.5$nm$, 748.1$nm$ and 893.1$nm$, respectively. 
  For each selected object we cut out a square region of roughly
  200''$\times$200'' (950 kpc at $z=0.34$), centered on the galaxy,
  from the five ugriz field images.
  We then shifted the resulting thumbnails using cubic convolution
  interpolation to center the main galaxy on the central pixel.
  Parts of the resulting thumbnails which extended beyond the SDSS
  stripe edge where given zero weight in the stacked images.

  \begin{figure}
    \includegraphics[width=0.47\textwidth]{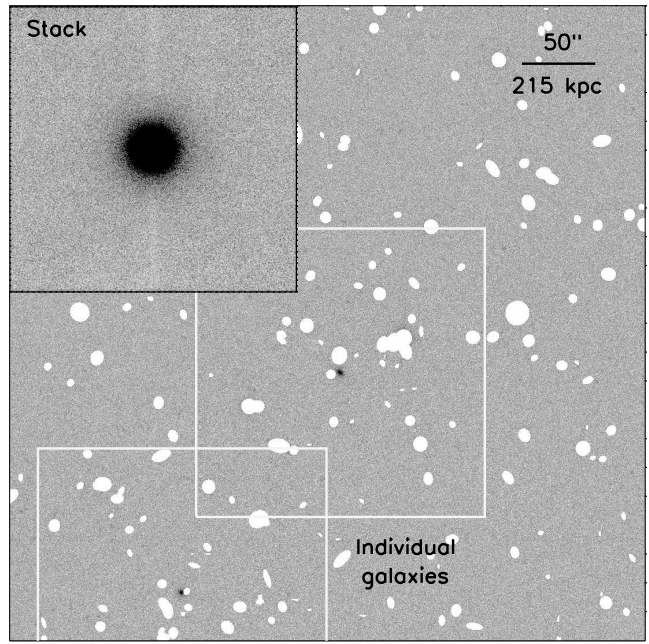}
    \caption{Image preparation for stacking: thumbnails of size
    200''x200'' pixels were cut around each LRG while all other objects in
    the field were masked out.  The stack in the top-left corner was
    made using more than 42550 LRG images and can be traced to a
    radius greater than 100 kpc.} 
    \hfill
    \label{fig:red_example}
  \end{figure}
  
  In order to detect and mask out any foreground and background
  objects we created a masking template by combining the thumbnails of
  three of the optical bands (r, i and z).
  This increased the signal-to-noise ratio of the template
  by a factor of roughly $\sqrt{3}$ compared to the individual frames,
  thus enabling us to unveil more sources in the field.
  We then ran SExtractor \citep{bertin_sextractor:_1996} on the
  combined image and detected all the objects in the frame.
  We set the detection threshold to a value of 1.4 times
  the standard deviation above the background RMS level and used AUTO
  photometry to extract Kron radii for the detected objects.
  Finally we constructed elliptical object masks by growing the
  semi-major and semi-minor axes as determined by Sextractor by a
  factor of 2.5.
  We found the optimal values for both the detection threshold and the
  mask growth factor from trial and error and verified that no
  additional unmasked objects remained in a visual inspection of
  the images.
  The resulting masks cover on average 18\% of the total area of each
  thumbnail.
  We note that galaxies below the detection threshold of SDSS are
  not excluded by our procedure.
  We return to this in subsection \ref{ranstack}.

 \subsection{Magnitude-bin stacks\label{magstack}}
  After creating object masks using the combined g+r+i images we
  continued on to preparing the individual band images for stacking.
  We did so by applying the master mask to the images in each of the
  bands and carefully subtracting both the soft bias, as introduced by
  the SDSS pipeline, and the remaining sky from the images.
  We determined the background sky level by temporarily masking out
  the central galaxy and fitting a Gauss curve to the pixel
  distribution of each masked thumbnail, assuming the residual sky
  pixel values are distributed like Poisson noise.
  This resulted in a set of five thumbnails per field with the
  central LRG being the only light source in each sky-subtracted
  frame.
  We compared the background level values we found to those obtained
  from the image headers, where those were available, and found
  excellent agreement.

  \begin{figure}
    \includegraphics[width=0.47\textwidth]{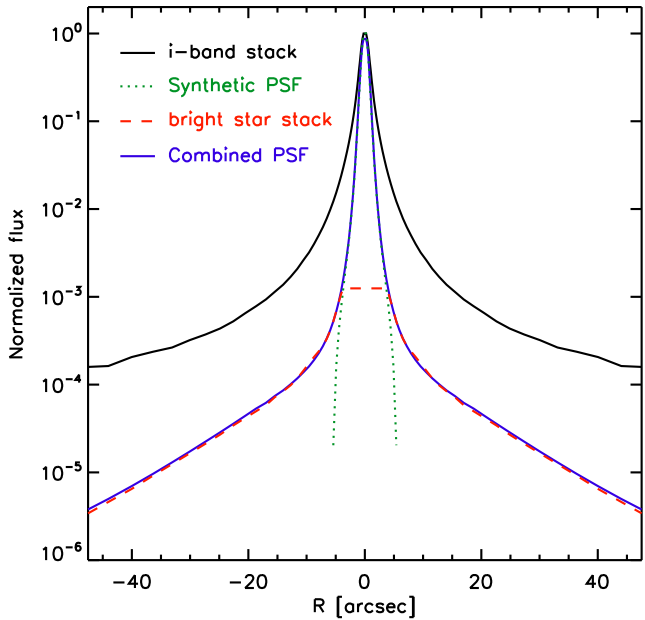}
    \caption{Radial light profile of the i-band stack (black solid
    line) overlaid with light profiles of the synthetic PSF stack
    (green dotted line), the bright star PSF stack (red dashed line) and
    the combined PSF (solid blue line).  Light scatter due to the PSF
    is most significant in i-band images.}
    \hfill
    \label{fig:psfoverlay}
  \end{figure}

  One of the challenging aspects of stacking images with any
  range of object brightnesses lies in the normalization of
  individual frames in respect to one another.
  To minimize normalization errors and optimize the S/N ratio we did
  not normalize each individual image but rather divided the sample
  into thirteen bins based on the apparent magnitude of the galaxies
  as determined by the SDSS pipeline.
  By doing so we ensured that the galaxies in each bin are
  ``pre-normalized''.
  The bins span a brightness range of 17 to 19.6 r-magnitudes at
  intervals of 0.2 magnitudes, corresponding to a flux difference of
  up to only 20\% within any given bin.
  These magnitude cuts were chosen such that only bins that had at
  least 100 galaxies in them will be processed.
  We note that we did not rescale the images to a common physical
  scale prior to stacking them.
  The $\pm$10\% variation in physical scale over the redshift range of
  the sample is likely small compared to the variation in effective
  radii between the individual galaxies.

  In addition to stacking masked, LRG centered images within each
  magnitude bin we summed up their corresponding object masks to serve
  as weight maps for the stacks.
  We then divided the summed images in each of the five bands by their
  respective weight mask, thus creating an averaged
  exposure-corrected stack.
  Since no normalization was required for images within the same
  magnitude bin, the noise characteristics were essentially improved
  by the square-root of the number of images in that bin.

 \subsection{Random Stacks\label{ranstack}}
  Correct sky subtraction is crucial for properly analyzing the faint
  outskirts of the galaxy stacks.
  However, since the residual ``sky'' is actually composed of at least
  three different light sources, the background subtraction performed
  in subsection \ref{magstack} has little physical meaning.
  These light sources include ``real'' sky background, light from
  undetected, and therefore unmasked, galaxies and - most importantly
  - light from the faint outer halo of the central LRG.
  In order to obtain a meaningful background measurement we stacked
  thumbnails at random locations within fields from the stacking list,
  using the same number of frames as are in the galaxy stacks and the
  exact same sky level values as determined for individual fields in
  the previous step. 
  For each of the optical bands we created one hundred random stacks
  in every magnitude bin to achieve a statistically meaningful
  distribution of the noise characteristics.
  We then averaged the random stacks and subtracted the resulting image
  from each of the magnitude-bin stacks.
  The background in the LRG stack is now defined as all emission in
  the vicinity of the LRG in excess of that of a random, nearby
  position.

 \subsection{Final steps\label{finstep}}
  As a last step prior to averaging the magnitude bin stacks
  we normalized them with respect to each other using the mean apparent 
  magnitude of their input images.
  We then summed the normalized stacks, weighting them by their
  relative total number of stacked frames.
  Finally, for the purpose of absolute photometry calibration we
  matched the final stack to the average apparent magnitude of the
  18.0$\leq$$m_r$$<$18.2 bin, resulting in five photometrically calibrated
  stacks in the u,g,r,i and z bands.
  We note that this method of normalizing magnitude binned stacks is
  significantly less sensitive to photometry errors than normalizing
  individual images.
  
  \begin{figure}[h]
    \begin{center}
      \includegraphics[width=0.4345\textwidth]{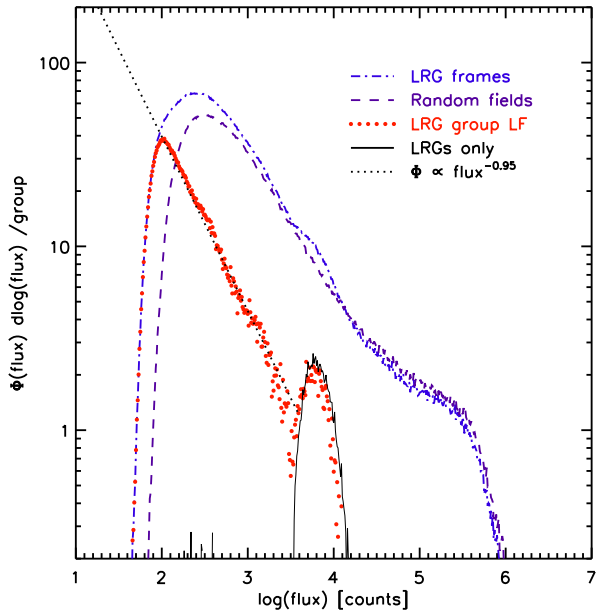}
      \caption{The empirical LRG group luminosity function (thick red
      dots) was derived by subtracting the luminosity function of
      objects in randomly selected frames (purple dashed line) from
      that of the LRG frames (blue dot-dashed line).
      The LRG only luminosity function is also plotted, along with the
      best fit power-law curve to the faint end slope.} 
      \hfill
      \label{fig:plot_lf}
    \end{center}
  \end{figure}
      
\section{Stack analysis}
 Each of the 42579 frames that went into making the final stack was
 obtained with 53.9 seconds of exposure time.
 The final stacks, then, have an effective depth that corresponds to an
 exposure time of roughly 2.3 Msec, equivalent to 40 hours on a 10m
 class telescope.
 Obviously, such deep images of individual objects would be
 completely dominated by light from neighboring objects, flat field
 errors and scattered light by the PSF.
 Stacking a large number of images may currently be the 
 best method for studying the faint outskirts of galaxies.
  
 \subsection{PSF stacks\label{psf_stacks}}
  When analyzing any deep dataset one must be aware of the effects
  of light scatter due to the point spread function in the images.
  Since the dynamic range in the data is large, the faint outskirts of
  the stack can in fact be dominated by scattered light from the
  bright galactic center.
  In order to determine the effects of the PSF on
  the galaxy stacks and to properly remove them we stacked synthesized
  PSF images created using Robert Lupton's Read Atlas Images
  code\footnote{http://www.sdss.org/dr7/products/images/read\_psf.html}.
  The PSF images were produced at the same CCD positions as the
  stacked LRG images for each of the observed bands.
  However, the synthetic PSFs only extend as far as $\sim$12'' (58
  kpc) in radius and do not reach the full extent of the LRG
  stacks, where PSF ``contamination'' may still be important
  \citep[see, for example,][]{de_jong_point_2008, bergvall_red_2009}.
  We therefore stacked all the bright star images from SDSS in the
  r-magnitude range 8.0$<m_r<$8.2 in each of the five optical bands.
  The resulting stacks are saturated inside of roughly 5'',
  where they are unusable, but extend out to a radius of 50''.
  We then combined the synthetic profiles with the bright
  star profiles to create PSF images in each band that include the
  effects of scattered light both at small and large radii.
  The synthetic, bright-star and combined PSF i-band profiles are
  shown in figure \ref{fig:psfoverlay} along with the stack light
  profile.  
  In the appendix we perform additional testing of the effects of
  different PSF models on our results.
 
 \subsection{PSF deconvolution and Surface brightness
    profiles\label{sur_bri}}
  Several techniques have been proposed and widely used for PSF
  deconvolution from imaging data \citep[e.g.][]{lucy_iterative_1974,
  hogbom_aperture_1974}.
  These algorithms, however, typically work well either in the inner
  parts of galaxies or their faint outskirts but not both
  simultaneously.
  We therefore chose to use the recently introduced technique of
  \cite{szomoru_confirmation_2010} for PSF deconvolution from our LRG
  stacks.
  Following this method, we first used Galfit 3.0
  \citep{peng_detailed_2002} to fit a 2D S\'{e}rsic model with the
  combined PSF images serving as the input convolution kernel.
  For the purpose of pixel weighting, to which the GALFIT results
  are sensitive, we supplied the software with the weight
  masks produced for each stack (subsection \ref{magstack}).
  We then constructed a synthetic S\'{e}rsic profile from the best-fit
  parameters to which we added the residuals that were left
  after subtracting the PSF convolved model from the image.
  In their paper, \cite{szomoru_confirmation_2010} showed that this
  method is insensitive to variations in the S\'{e}rsic parameters
  used to create the model.
  In order to increase the flexibility of our model fits we left the
  background level as a free parameter, in effect fitting the light
  profiles with a simple two-component model of a constant plus a
  S\'{e}rsic function.
  We also ran GALFIT with the background level set to zero in order to
  test how the lack of a constant component affects the model fits.
  Finally we measured the total flux in the light profile before and
  after deconvolving the PSF from it in order to verify that flux is
  globally conserved.
  We found that the difference between the total flux within the
  analysis radius of 475 kpc is less than 0.5\% between the
  deconvolved and the original profiles.

  Figure \ref{fig:rprofile} shows the PSF deconvolved profile of the
  r-band stack (the stack of galaxy images in the r-band) along with
  the GALFIT S\'{e}rsic model fit.
  To derive the profiles we first applied the IRAF task \textit{ellipse} 
  on the average of the r,i and z stacks while allowing the central
  position, ellipticity and position angle to vary with radius.
  We then used the output table from this fit as an input template for
  obtaining the light profiles of the stacks and models.
  The error bars were derived using randomly selected field stacks
  (see subsection \ref{erran} for details).
  Also shown in the figure are the initial r-band profile and the
  residual background level as determined by GALFIT.

  We note that an ideal deconvolution of the light profile would
  require a radially-varying PSF which is weighted by the number of
  galaxies stacked in each radius bin.
  However, we assume that this would have a negligible affect on the
  properties of the light profiles compared to other systematic
  effects and do not further modify the image PSF deconvolution.

 \subsection{Profile error analysis\label{erran}}
\begin{figure}
    \includegraphics[width=0.47\textwidth]{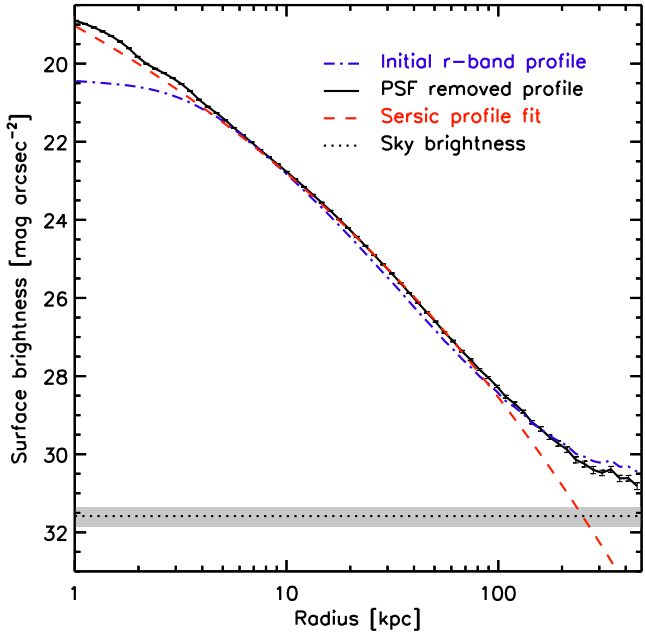}
    \caption{The PSF deconvolved profile of the r-band stack (black solid
    line) is shown along with the best GALFIT S\'{e}rsic model
    (red dashed line). The sky brightness level as determined by
    GALFIT is also plotted (black dotted line), along with the
    1$\sigma$ variation in the random stack measurements.
    The error bars are the standard deviation of a distribution of
    one hundred model fits using randomly selected stacked fields (see
    subsection \ref{erran} for details). This profile is corrected for
    undetected group galaxies using the empirical luminosity function
    described in subsection \ref{undet}}
    \hfill
    \label{fig:rprofile}
  \end{figure}

\begin{figure}
    \includegraphics[width=0.49\textwidth]{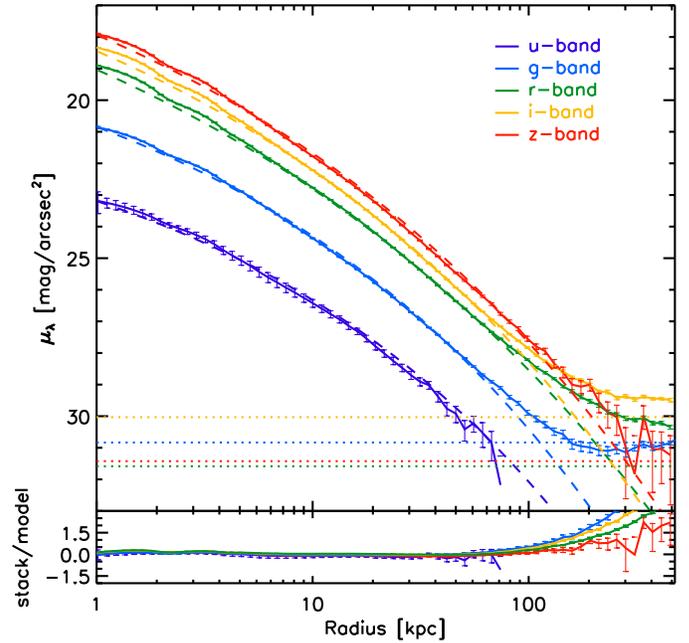}
    \caption{The light profiles of the five stacks are plotted along
    with their best fit model and the residual sky background as
    determined by GALFIT.
    The quotients of each model and light profile pair are also
    plotted, showing good agreement out to 100 kpc.
    The error bars are the standard deviation of a distribution of
    one hundred model fits using randomly selected stacked fields (see
    subsection \ref{erran} for details).}
    \hfill
    \label{fig:allprofs}
  \end{figure}

  The unprecedented depth and background uniformity of the stacks
  reveal faint emission hundreds of times dimmer than the typical LRG
  central surface brightness.
  In images of such low level emission statistical errors are not
  the only significant, and perhaps not even the main, source of
  noise.
  Errors in the faint outskirts of the stacks from undetected sources
  and remaining flattening issues, and their affect on the model fits
  become increasingly important but also increasingly hard to asses.
  We therefore used the same random stacks that are described in
  subsection \ref{ranstack} to measure the effects of such biases on
  the stack light profile properties.
  We subtracted each random background image from its corresponding
  magnitude bin stack and created one hundred stacks per filter,
  repeating the steps described in subsections \ref{magstack} and
  \ref{finstep}.
  We then followed the procedure described in subsection \ref{sur_bri}
  to deconvolve the PSF from the stacks and derived a
  S\'{e}rsic parameter set for each random stack.
  Table \ref{tab:serfits} shows that the scatter around the obtained
  S\'{e}rsic values due to these errors is small, suggesting that the
  fit is weighted toward the luminous inner part of the stacks.
  The S\'{e}rsic index 1$\sigma$ deviations of all but the faint u-band
  stack are under 0.1 and the effective radius 1$\sigma$ deviations
  are under 0.2 kpc.
  In addition, we used the IRAF task \textit{ellipse} to obtain radial
  light profiles for each of the random background subtracted stacks,
  using the same input template table as was used in subsection
  \ref{sur_bri}.
  The error bars in figures \ref{fig:rprofile}, \ref{fig:allprofs} and
  \ref{fig:pelet} reflect the 1$\sigma$ scatter of surface brightness
  profile fits to the random stacks.
  
  Systematic errors in the measured profiles that may rise from
  stacking images of galaxies with a range of S\'{e}rsic parameters
  seem to be of lesser significance.
  \citet[][appendix B]{van_dokkum_growth_2010} stacked hundreds of
  synthetic model images with randomly generated S\'{e}rsic profiles
  and showed that the stack effective radius and S\'{e}rsic index $n$
  match well with the average values of the stacked profiles.
  This suggests that no additional significant errors are introduced
  by our stacking technique.

 \subsection{Undetected light correction\label{undet}}
  Despite aggressive masking of foreground and background objects
  prior to stacking the data some light from undetected, and therefore
  unmasked, sources remains in the images.
  In order to correct the stack light profiles for this effect we used
  the LRG images to derive an empirical luminosity function for the
  LRG group environment at $z\sim$0.34.
  We started by using SExtractor to extract photometry for
  all the objects in the LRG frames using the detection limits that
  were used for object masking.
  We then binned the resultant values with a bin size of 0.01 dex in
  log space and produced a luminosity function (in practice we
  produced an observed brightness function).
  We repeated both steps for an identical number of randomly
  selected fields from the same SDSS imaging fields and produced a
  luminosity function for background and foreground sources.
  The LRG group luminosity function is then given by the difference
  between the number of sources in a particular brightness bin in the
  LRG fields and the number of random field sources in the same
  brightness bin (see figure \ref{fig:plot_lf}).

  The contribution of light from undetected objects to the stack
  profile comes from the faintest group members whose luminosity
  function is expected to be well fitted by a power-law function.
  We therefore divided the LRG group luminosity function into an LRG
  component and a power-law component (solid and dotted black curves
  in figure \ref{fig:plot_lf}).
  The best fit power law curve to the data is relatively shallow, with
  a slope of $\Phi \propto flux^{-0.95}$, implying that less
  than 0.3\% of the total unmasked light in the group (the total light
  in LRGs and undetected objects) comes from undetected sources.

  Finally, we extracted a radial light profile from the flux of
  resolved (and masked) objects in the LRG frames and fitted it
  with a single parameter S\'{e}rsic model.
  We then normalized the best fit curve such that the total flux under
  it equaled 0.3\% of the total LRG flux.
  Assuming that the radial light distribution of the unresolved
  sources is identical to that of the resolved objects we subtracted
  the resultant correction profile from the stack light profile.
  This correction did not change the LRG profile significantly and
  only moved it within the error bars for any given radius bin.
  
  \begin{figure}
    \includegraphics[width=0.47\textwidth]{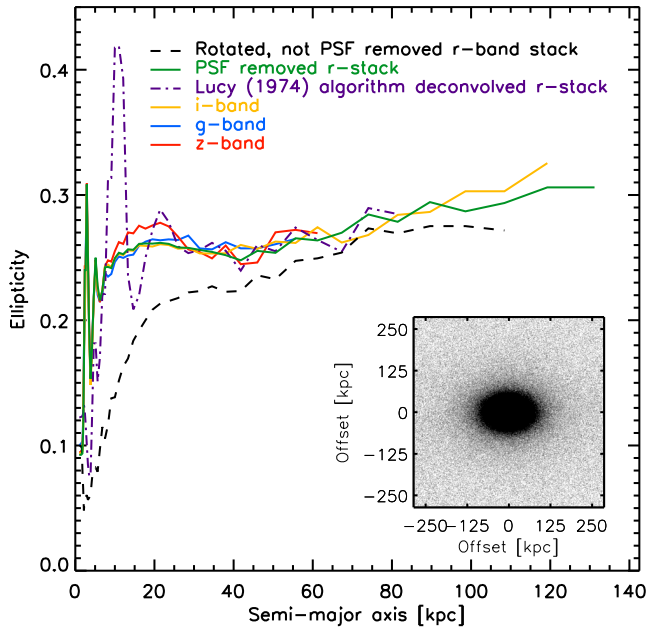}
    \caption{The ellipticity of the rotated stacks as a function of
    radius.  The solid curves represent the profiles of the PSF
    deconvolved stacks in the r, i, g and z bands.  The
    ellipticity in all filters is relatively flat out to roughly 60 kpc, 
    where a mild rise is observed.  
    The initial r-band stack profile (black dashed line) and the 
    \cite{lucy_iterative_1974} deconvolved profile (purple dot-dashed line)
    are also shown.
    The bottom-right inset shows the r-band rotated stack.}
    \hfill
    \label{fig:ellip}
  \end{figure}

 \subsection{S\'{e}rsic profile fits}
  The derived profiles of the g,r,i and z band stacks are well fitted
  by a single S\'{e}rsic parameter set out to $r$=100 kpc, at which
  radius the profiles deviate from the model by 0.2 mag arcsec$^{-2}$,
  corresponding to a difference of roughly 20\%.
  This is shown in figure \ref{fig:allprofs}, where the PSF-deconvolved
  profiles of all five stacks are plotted along with their best GALFIT
  models.   
  The determined S\'{e}rsic index values for the u and g band profiles
  are $n\sim$4 and are slightly shallower than those of the redder
  colors with $n\sim$5 for the r,i and z band stacks.
  The effective radii of all profiles except for the u-band stack are
  between 11 and 13 kpc with errors under 0.2 kpc.
  As might have been expected, setting the background level to zero in
  GALFIT increased the best fit effective radii and S\'{e}rsic
  parameters by up to 10\% as GALFIT attempted to fit the excess light
  at large radii.
  At radii larger than 100 kpc the profiles deviate from the model and
  excess light is observed in the g,r,i and z band stacks.
  Figure \ref{fig:allprofs} also shows that at $r>$ 200 kpc there is
  30\% to 70\% more light in the PSF-deconvolved profiles than in
  their respective best fit models.

 \begin{figure*}
    \includegraphics[width=0.98\textwidth]{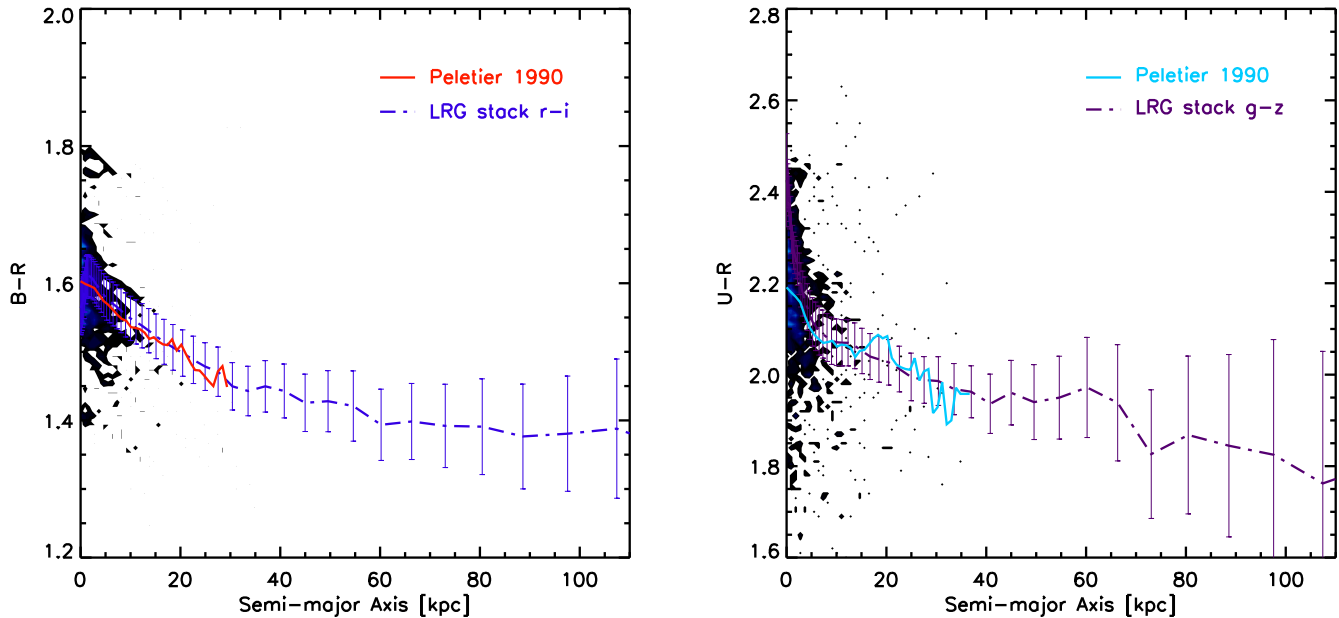}
    \caption{Broadband color comparison between the inner parts of
    nearby galaxies from \cite{peletier_ccd_1990} and the stacks.
    The blue density plots show the distribution of nearby galaxy
    colors out to about three effective radii and the red and light
    blue solid lines follow the running means of the sample.
    The blue and purple dot-dashed lines represent the r-i and g-z colors
    of the stacks, respectively, which match well with the
    \cite{peletier_ccd_1990} color profiles. Both
    colors are flat, within the error bars, outside of about 40 kpc.}
    \hfill
    \label{fig:pelet}
  \end{figure*}

  The derived profile parameters for the five final stacks are
  presented in table \ref{tab:serfits}.
   
 \subsection{Rotated LRG stacks}
  Early work in the field of massive galaxy evolution showed that
  the average ellipticity of nearby ellipticals is 0.3 $<\epsilon<$
  0.4, with a smaller value for the most massive systems
  \citep[e.g.,][]{sandage_intrinsic_1970, binney_apparent_1981}.
  The average ellipticity of LRGs can be measured from our image
  stacks. 
  We used SExtractor to derive a position angle for each of the 42579
  galaxies, rotated all frames to a common axis and stacked them
  following the steps described in section \ref{secstack}.
  We then deconvolved the PSF from the stacks using
  the technique discussed in subsection \ref{sur_bri}.
  Since this technique utilizes S\'{e}rsic parameter fits with
  radially constant ellipticities as the underlying model we tested
  our results by deconvolving the r-band stack using the
  \cite{lucy_iterative_1974} algorithm. 
  Figure \ref{fig:ellip} shows the rotated stack ellipticities in four
  of the bands, excluding the u-band stack due to its significantly
  noisier profile (as can be seen in figure \ref{fig:allprofs}).
  Also plotted are the original, not PSF deconvolved r-band stack (black
  dashed line) and the Lucy deconvolved r-stack profile (purple
  dot-dashed line).
  All profiles were obtained using the IRAF task \textit{ellipse} and all
  are flat out to roughly 60 kpc with an average 
  ellipticity value of 0.26 between 10 and 60 kpc ($\epsilon=0.21$
  when the centers are not excluded).
  The radial dependence of the ellipticity outside of 60 kpc appears
  to be mild, with a slight rise to $\epsilon\sim0.3$ at $r=100$ kpc.
  Furthermore, the profiles of both PSF-deconvolved and PSF-deconvolved
  stacks are in good agreement with each other, with less scatter in
  the inner parts of the former.
  From a similarly rotated stack of BCG images
  \cite{zibetti_intergalactic_2005} found a steep rise in profile 
  ellipticity outside of roughly 80 kpc followed by a steep decline at
  radii greater than 200 kpc.
  In the range 80 to 200 kpc this trend is similar to what we observe
  for LRGs, suggesting that there may be a continuum of properties
  going from group to cluster environments.
  We note that \cite{zibetti_intergalactic_2005} did not correct their
  stacks for PSF-induced effects.
  
  \begin{table}
    \caption{Stack S\'{e}rsic parameters}
    \centering
    \begin{tabular}{c c c}
      \hline\hline
      Filter & S\'{e}rsic index & Effective radius (kpc)\\
      \hline
      u & 3.94$\pm$1.62 & 17.0$\pm$9.80\\
      g & 4.03$\pm$0.09 & 12.6$\pm$0.19\\
      r & 5.50$\pm$0.05 & 13.1$\pm$0.10\\
      i & 4.86$\pm$0.05 & 10.9$\pm$0.06\\
      z & 4.91$\pm$0.08 & 11.5$\pm$0.12\\
      \hline\hline
      \label{tab:serfits}
    \end{tabular}
  \end{table}
  
 \subsection{Color gradients}
  It has long been known that the broadband colors of nearby
  elliptical galaxies vary with radius in the optical regime
  \citep{vader_three-color_1988, franx_multicolor_1989,
  peletier_ccd_1990}.
  Nevertheless, studies measuring such color gradients typically rely
  on observations of the inner parts of these objects, out to only 2-3
  effective radii.
  In figure \ref{fig:pelet} we plot the color profiles of r-i and g-z.
  The color pairs (r and i, g and z) have similar ellipticity profile
  sizes to ensure that the stack depths are comparable (figure
  \ref{fig:ellip}).
  We compare our results with a study of nearby ellipticals by
  \cite{peletier_ccd_1990} and show that the profile shapes of the two
  studies agree well out to a radius of $\sim$30 kpc.
  The stack color values are plotted in the observed frame and are
  matched to the \cite{peletier_ccd_1990} rest frame colors by adding a
  constant.
  We note that outside of about 20 kpc the local sample is composed of
  only a few galaxies where sufficient depth was achievable.
  With the advantage of our deep stacks, however, we are able to study
  the colors of LRGs out to more than eight effective radii.
  This is the first time that the colors of massive galaxies
  are observed at such large radius, providing a new insight on the
  stars that are found in the outskirts of massive galaxies.
  Indeed, the color profiles of LRGs change trend beyond the limit of
  nearby galaxy observations.
  Despite initially getting bluer in the inner 40 kpc of both plotted
  colors, the profiles quickly flatten out to a relatively constant
  value. 
  The outskirts of LRGs are then roughly 0.15 dex and 0.2 dex bluer
  than their centers in the $r-i$ and $g-z$ colors, respectively.

  The color profile is different from that observed by Z05 at r$>$20
  kpc as Z05 find that BCGs become very red at large radii.
  As we show in the appendix, the Z05 color profile may be severely
  affected by PSF effects at all radii, including the stack outer
  parts.
  Therefore, the different radial color profiles do not necessarily
  imply that LRGs and BCGs are fundamentally different objects.
  We note that the g-r color gradient found by Z05 has a similar slope
  as the g-z profile presented in figure \ref{fig:pelet}.
  The g-r profile of Z05 may suffer less from PSF effects than their
  r-i profile.

\section{Discussion}
 \subsection{Light profiles and the size of massive galaxies\label{sec:lp}}
\begin{figure}
    \includegraphics[width=0.49\textwidth]{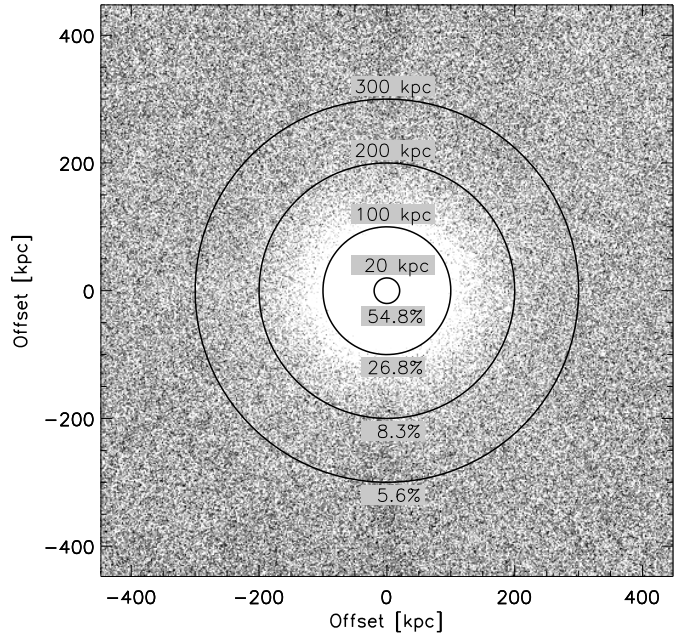}
    \caption{The radial stellar light fraction of the r-band stack in four
    radius bins of 20,100,200 and 300 kpc.  Note that a
    non-negligible fraction of the light is detected in the
    extreme outskirts of the stack.}
    \hfill
    \label{fig:lfrac}
  \end{figure}

  The first and foremost result that arises from this study is that
  faint, gravitationally bound stellar light can be traced in massive
  elliptical galaxies out to a radius of 100 kpc.
  By stacking a large number of faint galaxy images we detect light
  at such distance from the centers of massive galaxies with good
  confidence.
  In figure \ref{fig:lfrac} we show that the total accumulated light at
  20$< r/\rm kpc <$100 is non-negligible and accounts for roughly 27\%
  of the overall flux in the stack.
  In fact, more than 13\% of the stack light can be detected at very
  large radii outside of $r=$100 kpc, or more than 8 effective radii.
  This is especially interesting in light of recent studies that find
  compact massive galaxies at $z\sim$2, exhibiting effective radii
  of only 1 kpc \citep[e.g.,][]{daddi_passively_2005,
  trujillo_size_2006, van_dokkum_confirmation_2008}.
  The size growth of these objects is evidently rapid, expanding the
  physical scale of the galaxy by a factor of at least 5 in less than
  10 Gyr.
  Unfortunately, the physical growth mechanism cannot be directly
  observed in the LRG stacks as any signal from individual  galaxies
  is smoothed and averaged over the entire sample.
  Nevertheless, the lack of a clear change in the stack light profile
  slopes out to 100 kpc suggests that the observed light in the
  outskirts of LRGs is physically associated to the galaxies and their
  inner parts.
  Further evidence for this comes from the relatively
  radially-independent ellipticity profiles which vary only slightly
  out to 100 kpc.
  Any other light sources, such as background contamination or
  residual PSF scattering would be uncorrelated with the position
  angle of the LRG as measured in individual SDSS images, resulting in
  a circular light distribution.

  Outside of roughly 100 kpc the light profiles of the g,r,i and
  z-band stacks depart from the simple S\'{e}rsic model profile and
  exhibit excess light (figure \ref{fig:allprofs}).
  This departure from a simple model is observed here for the first
  time in LRGs and it shows that stars at the extreme outskirts of
  massive galaxies follow a different gravitational potential than
  stars in the inner parts.
  It is known that the potential at these radii is dominated by the
  properties of the dark matter halo, implying that the light profile
  is not necessarily expected to follow the same model that
  describes the inner stellar body.
  Alternatively, this excess light may simply be the residual
  background in the images, reflecting unresolved light from the group
  environment in which LRGs typically reside.

\begin{figure}
    \includegraphics[width=0.49\textwidth]{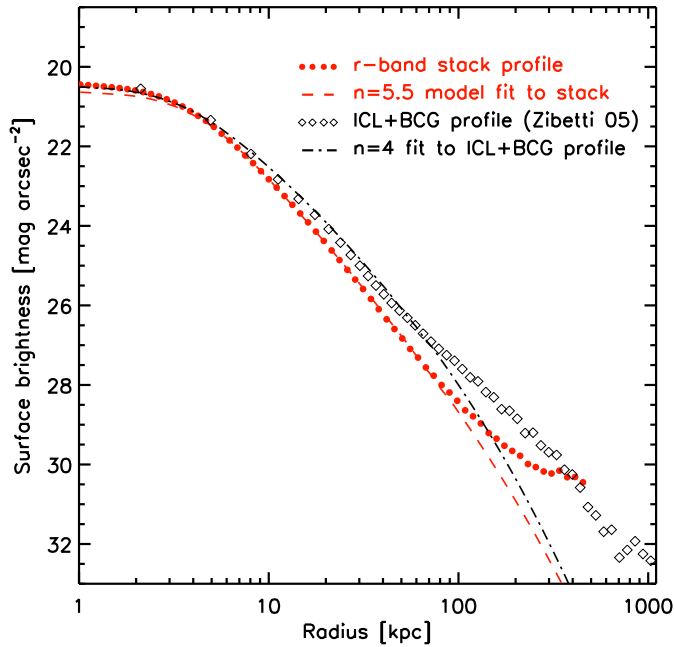}
    \caption{A comparison between the light profiles of our LRG r-band
      stack and the ICL profile from
      \cite{zibetti_intergalactic_2005}.
      The ICL profile departs from a single parameter S\'{e}rsic model
      at 50 kpc, or double the departure radius of 100 kpc that is
      observed in the LRG stack.
      This suggests a more significant population of intergalactic
      stars in massive clusters than in groups.}
    \hfill
    \label{fig:zibetti}
  \end{figure}

  Excess light was also observed by Z05, who studied the ICL around
  brightest cluster galaxies from a stack of 683 SDSS images.
  Such galaxies typically live in dense halos with total mass of
  $10^{14}$ to $10^{15} M\solar$ and are inherently different
  from LRGs whose group halos are a few times $10^{13} M\solar$ in
  mass.
  Z05 found that in clusters, this ``extra light'' constitutes
  only a small fraction of the total cluster profile, accounting
  for less than 11\% of the light inside of 500 kpc.
  Nevertheless, the ICL profile departs from a single parameter
  S\'{e}rsic model already at $r\sim50 kpc$, compared to the departure
  radius of 100 kpc that is observed in our LRG stacks
  (figure \ref{fig:zibetti}).
  This suggests that the massive clusters studied by Z05 may more
  readily support a population of intergalactic stars than the groups
  in which LRGs reside.
  In their paper Z05 correct their light profiles for unresolved
  cluster sources using the luminosity function given by
  \cite{mobasher_photometric_2003}.
  We note that the PSF, which is not deconvolved from the ICL+BCG
  profiles presented in Z05 may scatter light at all radii and
  increase the errors of the S\'{e}rsic model fit.

  Unlike the outer parts of LRGs, the centers of these galaxies are
  not well resolved in our stacks.
  Studies utilizing high resolution HST images showed that the profile
  at the inner parts of nearby ellipticals often departs from the
  S\'{e}rsic model that traces their outskirts.
  More specifically, the most massive ellipticals exhibit flattened
  central light profiles \citep[e.g.,][]{lauer_cores_1985,
  kormendy_hst_1994, lauer_centers_1995, faber_centers_1997}.
  Recently, \cite{kormendy_structure_2009} used a compilation of HST
  and ground based data to show that although well fitted
  by a S\'{e}rsic model out to large radii, the most massive Virgo
  ellipticals exhibit 1 kpc scale cores. 
  In our stacks we cannot resolve such physical scales as 1 pixel
  in the SDSS data is equivalent to 1.9 kpc at the stack mean
  redshift of 0.34.
  We are nevertheless able to confirm the excellent fit of massive 
  elliptical galaxy profiles to a single S\'{e}rsic profile out to a
  few effective radii that \cite{kormendy_structure_2009} found for 
  individual Virgo galaxies (reaching $\Delta\mu_\lambda\geq$ 0.2 mag 
  arcsec$^2$ at $r_\lambda\geq 100$ kpc).

 \subsection{How much light is missed?}
  The deep stacks allow us to test how much light is missed in typical 
  studies of the profiles of individual LRGs and derive a correction
  factor that can be applied in such cases.
  To do so we first selected all the LRGs in a single magnitude bin, 
  18.0$\leq$$m_r$$<$18.2, and used GALFIT to produce a S\'{e}rsic model
  to each object individually.
  We then excluded all fits with errors of more than 10\% in either
  the n parameter or the effective radius, resulting in a
  mean effective radius value of 11.7 kpc.
  The difference between this value and the one derived by GALFIT
  for the stacked image ($r_e=$13.1 kpc) is then $\sim$10\%.
  This implies that surveys may underestimate the size
  of massive red galaxies by this amount.
  The total flux in the stack, however, accounts for $\sim$20\% more
  than the mean value for the individually derived profiles,
  suggesting that a non-negligible amount of light is typically missed
  and that the total stellar mass is underestimated.

 \subsection{Minor mergers and the LRG color profile}
  It has long been known that the color profiles of nearby massive
  ellipticals exhibit a relatively smooth gradient toward bluer
  colors from the galaxy centers outward.
  Line index measurements \citep[e.g.,][]{carollo_metallicity_1993,
  davies_line-strength_1993,spolaor_early-type_2010} and studies of
  high redshift galaxies \citep{tamura_origin_2000} provide evidence
  that this observed color gradient originates from a radial slope in
  stellar metallicity.
  It has also been claimed in these studies that age plays only a
  secondary role in producing this trend.
  Nevertheless, it is yet to be determined which physical mechanisms
  create and evolve the observed color gradient.
  With the color profile in mind we will discuss in this section a
  previously suggested model in which minor mergers and low mass
  accretion events are the main supplier of stars to the outskirts of
  massive galaxies.
  We will also utilize our LRG stacks to provide additional pieces of
  evidence that support this model.

  \subsubsection{Creating the faint stellar halo}  
   It has been suggested that the size and mass growth of massive
   galaxies are dominated by minor mergers which contribute a
   relatively constant flow of accreted mass over a long time.
   This is evident from analytic calculations
   \citep[e.g.,][]{bezanson_relation_2009}, numerical simulations of
   large volumes \citep[e.g.,][]{de_lucia_hierarchical_2007,
     naab_formation_2007} and observations of nearby elliptical galaxies
   \citep[e.g.,][]{van_dokkum_recent_2005, tal_frequency_2009}.
   The contribution of accreted mass, however, is not expected to
   be uniform throughout the galaxy.
   This can be shown using very simple energy arguments following the
   assumption that accreted mass is more likely to stay near the
   radius at which is was accreted if its total energy budget is
   similar to the gravitational energy of the accreting galaxy at that
   radius.
   In other words, if the escape velocity of a star in an infalling
   galaxy equals the orbital velocity of the accreting galaxy at the
   radius at which the star escapes, this star will likely fall into
   orbits close to that radius.
   If on the other hand the star escapes the infalling galaxy at any
   other radius it will either fall closer to or be thrown farther
   from the center of the accreting galaxy, thus contributing to a
   more uniform mass distribution.
   This argument can be described by the following equation:
   \begin{equation}
     V_{circ}^M = V_{esc}^m
   \end{equation}
   where $V_{circ}^M$ is the circular velocity of the accreting galaxy
   and $V_{esc}^m$ is the escape velocity from the infalling galaxy.
   This can be expressed in terms of the masses of the accreting and
   infalling galaxies, $M$ and $m$ and the accretion and escape radii,
   $R_a$ and $r$,
   \begin{equation}
     R_a = \frac{M}{m}\frac{r}{2} \label{eqn:r_acc}
   \end{equation}
   Equation \ref{eqn:r_acc} implies that more massive infalling
   galaxies will preferentially deploy most of their stars closer to
   the center of the accreting galaxy.
   However, the mass ratio $M/m$ cannot be close to 1 since such major
   merger will violently disrupt both galaxies and not simply disperse
   the stars of one into the stellar body of the other.
   In addition, if the mass ratio is high, the contribution of the
   infalling galaxy to the colors of the accreting galaxy will be
   minor.
   We therefore assume typical values of 4$<$$M/m$$<$10 for the
   mass ratio and 2$<r/$kpc$<$10 for the escape radius, both roughly
   correspond to the accretion of an $L^{\star}$ type galaxy.
   It is therefore convenient to re-write equation \ref{eqn:r_acc}
   using the suggested average values for $M/m$ and $r$:
   \begin{equation}
     R = 42\left(\frac{M/m}{7}\right)\left(\frac{r}{6\rm
     kpc}\right)\rm kpc
   \end{equation}

   This simple model provides a possible explanation for the observed
   size growth from $z$=2, as well as the smooth gradient and
   flattening in the stack color profile.
   It enables infalling low mass galaxies to easily deploy the
   majority of their stars outside of the accretion radius $R_a$ and
   it suggests that scattered stars that are stripped from their
   galaxy at any other radius may end up close to the center. 
   This may be the case if the accreting galaxy is indeed red and if a
   significant part of all infalling galaxies are bluer than the LRG.
   In addition, this scenario supports minor mergers with large $M/m$
   values as the main size growth mechanism as stars are preferentially
   being deployed far from the center of the accreting galaxy without
   increasing its total mass significantly.

  \begin{figure}
    \includegraphics[width=0.49\textwidth]{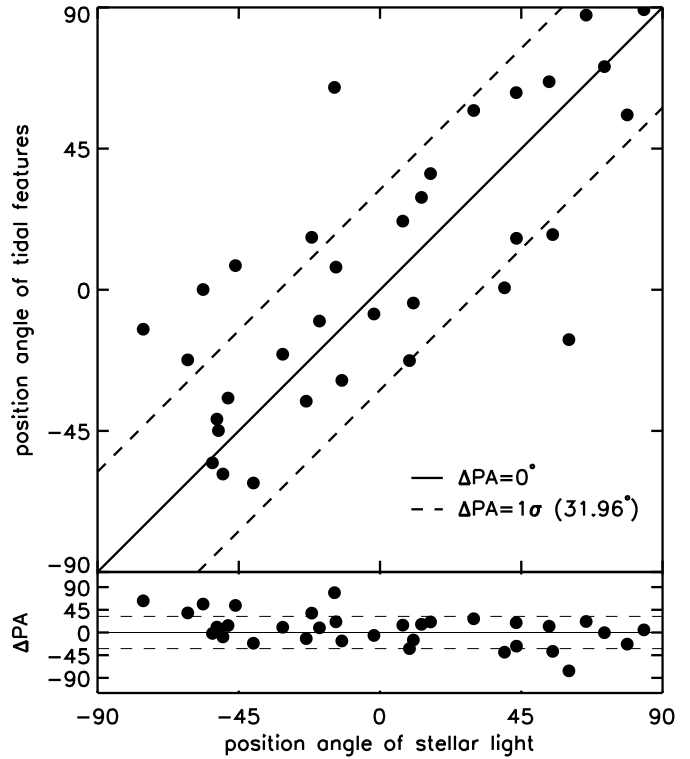}
    \caption{A comparison between the position angle of stellar
    light and that of tidal feature light in galaxies from the OBEY
    survey \citep{tal_frequency_2009}.
    The correlation between the orientation of tidal features and
    the stellar body suggests that minor interactions cannot be ruled out 
    as a source of stellar light to the outskirts of massive galaxies.}
    \hfill
    \label{fig:ellcomp}
  \end{figure}

  \subsubsection{The frequency of minor mergers}
   The frequency of low mass accretion events is a critical factor in
   assessing the importance of these interactions to the size and mass
   growth of massive ellipticals.
   Minor mergers must not be rare in order for the compact galaxy at
   $z$=2 to experience the observed rapid size growth.
   We now also know that the light at $r>$ 20 kpc comprises roughly
   40\% of the total stellar light (figure \ref{fig:lfrac}),
   suggesting that a constant stream of accreted mass is required.
   This implied mass growth confirms the results from
   \cite{van_dokkum_growth_2010} who found that nearby ellipticals are
   roughly twice as massive as compact $z$=2 galaxies.

   Observations of early type galaxies reveal ample evidence
   for close gravitational interactions between ellipticals and their
   neighboring galaxies.
   Such interactions, mainly in the form of minor mergers, leave their
   signature on the stellar bodies of the massive galaxy for a time
   period that depends on the interaction itself and that can last for
   up to a few Gyrs.
   Initial attempts to characterize tidal features around nearby
   ellipticals relied heavily on qualitative descriptions of the
   remnant morphology but nevertheless revealed evidence for
   interaction in many systems \citep{schweizer_correlations_1992}.
   More recent studies quantified an interaction parameter by
   measuring excess light in the tidal features compared to smooth
   models of the systems.
   Such studies analyzed volume limited samples of nearby galaxies,
   finding that more than 70\% of all ellipticals have had one or more
   recent minor merger events \citep{van_dokkum_recent_2005,
     tal_frequency_2009}.
   The mass growth that is inferred from this high interaction
   rate can be estimated at a factor of 2-3 since $z$=2 (see
   \citealt{tal_frequency_2009} for details).

  \subsubsection{Minor mergers and the ellipticity profile}
   Assuming that the distribution of infalling galaxies and
   therefore also tidal features is spherically symmetric around the
   elliptical, the stack outskirts are expected to be relatively
   round, in apparent conflict with the trend shown in figure
   \ref{fig:ellip}.
   It is however possible that gravitational interactions contribute
   to the ellipticity of the host galaxy or alternatively occur
   preferentially along the major axis of the dark matter halo, thus
   aligning the resultant tidal features with the position angle of
   the elliptical.
   In order to test this we examined galaxies from the
   OBEY survey \citep{tal_frequency_2009}, a complete sample of nearby
   ellipticals that was utilized to study and quantify tidal features
   in the stellar bodies of these objects.
   We measured the overall ellipticity of each galaxy with a tidal
   parameter value of at least 0.07 and compared it to the second
   moment of light distribution in the model-subtracted residual
   image.
   By measuring the second moment of residual light distribution we
   quantified, in effect, the position angle of tidal features.
   Figure \ref{fig:ellcomp} shows that a correlation indeed
   exists between the orientation of the stellar body and that of the
   tidal features. 
   A Spearman's rank test finds that the probability that this correlation
   was drawn from a random distribution is less than 0.1\%
   We also note that only two galaxies exhibit $\Delta PA>60^\circ$.
   Although such a suggestive correlation does not strongly support
   any single scenario as the main mechanism for creating the observed
   faint halos, its existence hints that minor mergers at least
   cannot be ruled out as a significant contributor to the stellar
   bodies of LRGs at large radii.

  This analysis is therefore consistent with the idea that minor
  mergers and gravitational interactions likely play an important role
  in determining the properties of massive red galaxies.
  Foremost is the blue color index of the profile at $r>40$ kpc which
  suggests that the outer halo is composed of younger or alternatively
  more metal poor stars compared to the center.
  This probably means that the stellar populations of the outskirts
  were formed separately from those in the center and probably
  accreted at a later time.
  This scenario is supported by the observed high rate of tidal
  features around nearby elliptical galaxies, although we note that
  the LRGs are typically more massive by a factor of $\gtrsim$2 than
  the galaxies in the \cite{tal_frequency_2009} sample.
  Our analysis shows that such accretion events likely deploy most of
  the stars at large radii.

\section{Summary and conclusions}
 In this paper we stacked more than 42000 images of luminous red
 galaxies in order to study the faint light of these objects at large
 radii.
 In our stacks we detected stellar light out to radii greater than 100
 kpc, thus providing a correction factor to the true size and overall
 stellar mass of LRGs.
 This is the first time that such an analysis is performed using a
 dataset of this scale, reaching unprecedented depth for $z>0.01$
 galaxies.
 The relatively flat ellipticity profile (figure \ref{fig:ellip})
 verifies that the light detected at large radii is physically
 associated with the stellar body.
 Interestingly, the profiles suggest increased ellipticity at large
 radii of the average LRG profile.
 
 In agreement with \cite{kormendy_structure_2009}, we confirmed that
 on average, the light profiles of massive ellipticals can be well
 described by a single S\'{e}rsic model out to roughly eight
 effective radii.
 Outside of 100 kpc the profiles deviate from a simple S\'{e}rsic
 model and exhibit extra light in the r, i and z-band stacks.
 This excess light can probably be attributed to unresolved intragroup
 light or a change in the light profile itself.
 Differentiating between these possibilities, however, may be
 difficult as both can have a similar effect on the profile shape at
 large radii.

 Finally, we utilized the five optical bands of SDSS to study the
 colors of these galaxies and showed that the well known decrease in
 color index out to 2-3 effective radii flattens out and stays blue
 compared to the galactic centers out to the detected stack limit.
 Although this finding by itself does not favor any one stellar
 population evolution scenario, it suggests that the central 20 kpc
 evolve somewhat differently from the rest of the galaxy.
 Previous studies of line indices in early type galaxies suggest that
 this difference can probably be attributed to a difference in
 metallicity.

 \begin{acknowledgements}
   We are grateful to Marijn Franx for many useful comments and to
   David Wake and Adam Muzzin for numerous engaging discussions.
   We also thank Stefano Zibetti for sharing his ICL light profiles
   with us. 

   Funding for the SDSS and SDSS-II was provided by the Alfred
   P. Sloan Foundation, the Participating Institutions, the National
   Science Foundation, the U.S. Department of Energy, the National
   Aeronautics and Space Administration, the Japanese Monbukagakusho,
   the Max Planck Society, and the Higher Education Funding Council
   for England. The SDSS was managed by the Astrophysical Research
   Consortium for the Participating Institutions.

   Funding for the SDSS and SDSS-II has been provided by the Alfred
   P. Sloan Foundation, the Participating Institutions, the National
   Science Foundation, the U.S. Department of Energy, the National
   Aeronautics and Space Administration, the Japanese Monbukagakusho,
   the Max Planck Society, and the Higher Education Funding Council
   for England. The SDSS Web Site is http://www.sdss.org/.

   The SDSS is managed by the Astrophysical Research Consortium for
   the Participating Institutions. The Participating Institutions are
   the American Museum of Natural History, Astrophysical Institute
   Potsdam, University of Basel, University of Cambridge, Case Western
   Reserve University, University of Chicago, Drexel University,
   Fermilab, the Institute for Advanced Study, the Japan Participation
   Group, Johns Hopkins University, the Joint Institute for Nuclear
   Astrophysics, the Kavli Institute for Particle Astrophysics and
   Cosmology, the Korean Scientist Group, the Chinese Academy of
   Sciences (LAMOST), Los Alamos National Laboratory, the
   Max-Planck-Institute for Astronomy (MPIA), the Max-Planck-Institute
   for Astrophysics (MPA), New Mexico State University, Ohio State
   University, University of Pittsburgh, University of Portsmouth,
   Princeton University, the United States Naval Observatory, and the
   University of Washington.

 \end{acknowledgements}
 
\bibliography{ms}
\bibliographystyle{apj}

\newpage

\appendix \label{app:psf}
 \section{The wings of the PSF profile}
  Light scattered by the wings of the PSF profile was shown to
  contribute noticeably to the color profiles of the outskirts of
  spiral galaxies.
  In his paper, \cite{de_jong_point_2008} argues that the choice of
  PSF model used for image deconvolution is crucial for analyzing data
  of spiral galaxies.
  The author further claims that the red stellar halos found by
  previous studies were artificially produced by PSF models that were
  poorly sampled at large radii.
  In this appendix we confirm this observation and show that the widely
  used synthetic PSF images may lead to inaccurate color gradients
  even in the peaky light profiles of massive red galaxies.

\begin{figure}[h]
    \begin{center}
      \includegraphics[width=0.49\textwidth]{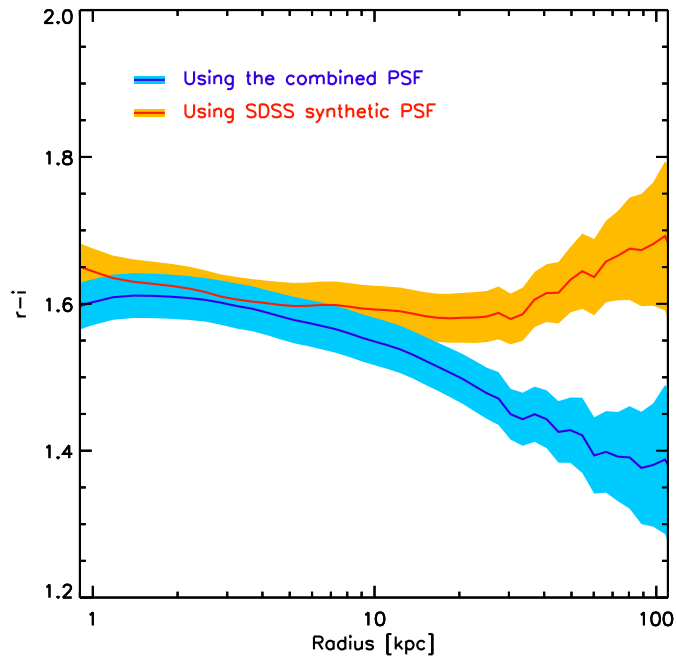}
      \caption{Color profile comparison between the combined PSF
	deconvolved stack (blue line) and the synthetic only PSF deconvolved
	stack (red line).
	The shaded areas show the 1$\sigma$ error bars derived using
	randomly selected field stacks (subsection \ref{erran}).
	The use of insufficiently sampled PSF profile artificially creates
	an red halo in our stacks.}
      \hfill
      \label{fig:psf_comp}
    \end{center}
  \end{figure}

  In order to test the effects of PSF profile selection on the
  properties of the derived S\'{e}rsic model we followed the
  procedure described in subsection \ref{sur_bri} to deconvolve the
  stacks using two different PSF images in each observed band. 
  The first profile is identical to the one used throughout this study
  and it combines a synthetically produced PSF model with a bright
  star stack (subsection \ref{psf_stacks}), while the second utilizes
  only the synthetic PSF image produced by the Read Atlas Image code.
  While the u,g,r and z models are less than 15\% larger when using
  only the synthetic PSF image, the i-band stack exhibits an increase
  of more than 25\% in this case.
  The increased growth in model size in the i-band compared to the
  other stacks implies that the halo of these galaxies may erroneously
  appear to be red.
  This is most clearly evident in figure \ref{fig:psf_comp}, where we plot
  the r-i color profiles of the two PSF deconvolved models and show that
  in the synthetic only case an artificial red halo appears outside of
  roughly 40 kpc.
  We conclude, in agreement with \cite{de_jong_point_2008}, that a
  proper choice of well sampled PSF model is critical for studying the
  faint outskirts of galaxies.
  
\end{document}